\documentclass[superscriptaddress,aps,amsfonts,notitlepage,onecolumn,nofootinbib]{revtex4-2}

\usepackage{xcolor,graphicx}
\usepackage{breqn,amsmath,amssymb}
\usepackage{mathtools}
\usepackage{enumerate}

\usepackage[russian, english]{babel}

\usepackage{graphicx}
\graphicspath{{Figures/}}
\usepackage{float}
\usepackage{subfigure}
\usepackage{appendix}
\usepackage{epstopdf}
\usepackage{dblfloatfix}
\usepackage{morefloats}
\usepackage[skip=5pt]{caption}

\usepackage{booktabs}
\usepackage{makecell}

\usepackage{comment}

\usepackage{subcaption}

\usepackage[T1]{fontenc}
\usepackage{natbib,hyperref}

\usepackage{mathtools}

\usepackage{centernot}

\usepackage[utf8]{inputenc}

\definecolor{linkcolor}{HTML}{799B03}
\definecolor{urlcolor}{HTML}{799B03}
\usepackage[english]{babel}
\usepackage{amsmath, amsthm, amsfonts, amssymb,amscd}
\usepackage[all]{xy}
\usepackage{graphicx}
\usepackage{import}
\usepackage{xifthen}
\usepackage{float}
\usepackage{transparent}
\usepackage{multirow}
\usepackage{cellspace}
\usepackage{colortbl}
\usepackage{bbm}
\usepackage{mathrsfs}
\usepackage{csquotes}
\usepackage{natbib}

\graphicspath{{pic/}}
\def\[{\begin{equation}}
\def\]{\end{equation}}

\DeclareUnicodeCharacter{0301}{`}

\makeatletter
\let\cat@comma@active\@empty
\makeatother

\usepackage{IEEEtrantools}

\begin{document}

\title{Inflation after Curvature Bounce}


\author{Rinat Kagirov}
\email{kagirov.rr19@physics.msu.ru}
\affiliation{Institute for Nuclear Research of the Russian Academy of Sciences, 
60th October Anniversary Prospect, 7a, 117312 Moscow, Russia}
\affiliation{Department of Particle Physics and Cosmology, Physics Faculty, M.V. Lomonosov Moscow State University,
Vorobjevy Gory, 119991 Moscow, Russia}

\begin{abstract}
We present a stable cosmological model of a closed universe in the presence of conventional scalar field. The stability of the model and the absence of singularity is ensured by spatial curvature without the need for additional peculiar matter. We reconstruct the Lagrangian and numerically compute observational predictions, including the number of e-folds, the spectral index $n_s$, and the tensor-to-scalar ratio. We present several sets of parameters that satisfy the current observational data.
\end{abstract}

\maketitle

\section{Introduction}

Observations indicate that the universe is nearly flat, but they cannot rule out a closed universe.
The role of a positive curvature is negligible at late times, but can be crucial in the early universe.
In particular, positive curvature allows for cosmologies that originate as Einstein static universe \cite{Ellis_2003, ORaifeartaigh:2017uct} or bounce cosmology \cite{ 
Murphy:1973zz,  Starobinsky, Durrer:1995mz, Falciano:2008gt, NOVELLO_2008, Battefeld_2015},
inflate, and later reheat to a hot big bang era. These cosmologies have no singularity, no
'beginning of time', and no horizon problem. If the initial radius (bounce radius) is chosen to be above the Planck
scale, they also have no quantum gravity era, so such models can be considered classically.

Models of inflation in a closed universe have been proposed before, e.g. model on potential in GR \cite{2Ellis_2003, massivebaounceRenevey_2021}. More recently, many different cosmological models, such as Genesis or bounce \cite{ Creminelli:2010ba, Easson:2011zy, Qiu_2011, Cai_2012, Osipov:2013ssa,  Kolevatov:2017voe, Cai_201722, Volkova2019, Mironov_2020, Mironov_2023}, have been proposed in the framework of Horndeski theories \cite{horndeski1974second} and its generalizations. As we are interested in a closed universe, we consider a generalization of Horndeski theory to the case with non-zero curvature \cite{Akama:2017jsa}.

The principal characteristic of our model is the absence of gradient and tachyon instabilities,  meanwhile there is no NEC(Null Energy Condition)-violating matter throughout its evolution \cite{Rubakov_2014}, facilitated by the presence of non-zero curvature. Furthermore, the model eliminates the initial singularity, thereby rendering the universe's evolution analogous to a bounce scenario within a closed universe framework. This curvature permits circumvention of the singularity theorems' conditions, leading to a transition from a contraction to an inflationary era. Consequently, curvature serves to avoid both instabilities and initial singularity, but without NEC violation.

One of the distinctions from previously proposed models lies in the consideration of the k-essence subclass of Horndeski theories, which in our case can be reduced by a field redefinition to a scalar field with a nontrivial potential and a standard kinetic term. In this case, the action takes the form of General Relativity (GR) with an additional conventional scalar field, which is appropriate to construct different single-filed cosmology scenarios. The idea to use spatial curvature to avoid singularity is very old, but up to day inspires a lot of interesting works (see \cite{Matsui_2019, Daniel_2023} for example). We too consider a simple type of Lagrangian for a scalar field in a homogeneous Universe, but ideologically we do a backward study. We analytically reconstruct the Lagrangian for an explicit behavior of the scale factor. Additionally we numerically calculate the parameters of the spectrum.

In this way, we introduce a set of parameters in the scale factor that retain all the key features of our model while influencing observable data, such as the r-ratio, spectral index $n_s$ and the number of e-foldings. This allows us to construct a stable model with adjustable parameters, enabling adaptation to current or future observations.

This paper is organized as follows. In Section \ref{2}, we provide the basic background on k-essence theory in a closed universe. In Section \ref{3}, we introduce the way to reconstruct the Lagrangian, present our model, and discuss the BKL instability as well as the absence of fine-tuning. In Section \ref{4}, we compare the predictions of our model with current observations. We conclude in Section \ref{5}.

\section{ k-Essence theory in open
and closed universe}\label{Horndeski in open
}
\label{2}

We examine cosmology in the $\mathcal{L}_{2}$ subclass of Horndeski theory, commonly referred to as k-essence \cite{Babichev_2008,  Deffayet:2011gz}, with non-zero curvature \cite{Akama:2017jsa}, specified by the following action:

\begin{align}\label{action}
S=\int d^4x\sqrt{-g}\,{\cal L}_{\rm k},
\end{align}
with
\begin{align}\label{lagrangian}
\mathcal{L}_{\rm k}&=G_2(\phi,X)+G_4(\phi)R,
\end{align}
where $X:=-g^{\mu\nu}\nabla_\mu\phi\nabla_\nu\phi/2$ and
we denote $\partial G/\partial X$ by $G_{X}$.

We use FLRW metric with spatial curvature,
\begin{align}
ds^2=-N^2(t)dt^2+a^2(t)\gamma_{ij}dx^idx^j,\label{metricansatz}
\end{align}
where $\gamma_{ij}$ is the metric of
maximally symmetric spatial hypersurfaces.
It can be written explicitly as
$\gamma_{ij}dx^idx^j = d\chi^2+S^2_{\cal K}(\chi)d\Omega^2$
with
\begin{align}
S_{\cal K}(\chi):=
\begin{cases}
\sin(\sqrt{\mathcal{K}}\chi)/\sqrt{\mathcal{K}}
 &({\rm closed}:\;{\cal K}>0),\\
\chi  &({\rm flat}:\;{\cal K}=0) ,\\
\sinh(\sqrt{-\mathcal{K}}\chi)/\sqrt{-\mathcal{K}}
& ({\rm open}:\;{\cal K}<0) ,\\
\end{cases}
\end{align}
and $d \Omega^2 := d\theta^2+\sin^2\theta d\varphi^2$.

We substitute metric to the action,
vary it with respect to $N(t)$ and $a(t)$, take $N=1$, and obtain the background equations of motion (Friedmann equations)  in the following form,

\begin{align}
2XG_{2X}-G_2-6H^2G_4-3\mathcal{G_T}\frac{\mathcal{K}}{a^2}&=0, \label{eom1}
\\
G_2+2(3H^2+2\dot{H})G_4+\mathcal{F_T}\frac{\mathcal{K}}{a^2}&=0, \label{eom2}
\end{align}
where $\mathcal{G_T}$ and  $\mathcal{F_T}$ are coefficients from second order action for tensor perturbations, which are both equal to 1 in our subclass. Further we assume $G_4(\phi)=\frac{1}{2}$ ($M_{pl} =1$), so the only non-trivial term is $G_2(\phi,X)$,  which gives the most general k-essence.

It is important to mention that the effect of curvature becomes negligible for large values of $a$, which is already true after several e-folds.

Below we study the quadratic action for scalar and tensor perturbations which is necessary to obtain Power spectra.

\subsection{Scalar perturbations}

Let us first consider the Lagrangian of scalar perturbations. After integration out the constraints \cite{Kobayashi:2019hrl} quadratic Lagrangian in the scalar sector reduces to one degree of freedom.

\begin{equation}\label{actionscal}
S_\zeta^{(2)}=\int \mathrm{d} t \mathrm{~d}^3 x a^3\left[\mathcal{G}_S \dot{\zeta}^2-\frac{\mathcal{F}_S}{a^2}(\partial \zeta)^2\right]
\end{equation}

where $\mathcal{F}_S, \mathcal{G}_S$ in our subclass are as follows:

\begin{equation}
    \mathcal{F}_S=-\frac{\dot{H}}{H^2}+\frac{\mathcal{K}}{H^2 a^2},
\end{equation}

\begin{equation}
    \mathcal{G}_S=\frac{X G_{2X}+2X^2G_{2XX}}{H^2}.
\end{equation}

Using equations of motion \eqref{eom1} and \eqref{eom2} it is easy to obtain $\mathcal{F}_S$ in this form: 

\begin{equation}
    \mathcal{F}_S=\frac{X G_{2X}}{H^2} ,
\end{equation}

so we can get expression for the speed of scalar perturbation:

\begin{equation}\label{cs}
    c^2_S=\frac{\mathcal{F}_S}{\mathcal{G}_S}=\frac{G_{2X}}{G_{2X}+2XG_{2XX}}. 
\end{equation}

Note, that there is no explicit dependence on the curvature in this form. But due to $\mathcal{K}$ term in the background EOM the dynamics of perturbation is different in flat, close and open Universe. 

The form of action \eqref{actionscal} dictates stability conditions,

\begin{equation}\label{stable2}
    \mathcal{G}_S \ge \mathcal{F}_S > \epsilon>0,  
\end{equation}

and equation of motion for the Fourier modes:

\begin{equation}\label{EOMscal}
\frac{1}{a^3 \mathcal{G}_S} \frac{\mathrm{d}}{\mathrm{d} t}\left(a^3 \mathcal{G}_S \frac{\mathrm{d} \zeta}{\mathrm{d} t}\right)+\frac{k^2 c_S^2}{a^2}  \zeta=0
\end{equation}

For a given Lagrangian of the theory, this equation can be solved numerically with vacuum initial conditions for canonically normalized scalar perturbations:

\begin{equation}\label{ICscal}
    \zeta(k, -\infty)=\frac{1}{a(t) (4\mathcal{G}_S \mathcal{F}_S)^{\frac{1}{4}}}\frac{\exp{-i k \eta(t)}}{\sqrt{2c_sk}},
\end{equation}

where conformal time is defined as

\begin{equation}
    \eta(t)=\int_{-\infty}^{t}\frac{\mathrm{d} t'}{a(t')}.
\end{equation}

\subsection{Tensor perturbations}

The same analysis can be performed for tensor perturbations \cite{Kobayashi:2019hrl}. Starting from the action:

\begin{equation}
S_T^{(2)}=\frac{1}{8} \int \mathrm{d} t \mathrm{~d}^3 x a^3\left[\mathcal{G}_T \dot{h}_{i j}^2-\frac{\mathcal{F}_T}{a^2}\left(\partial h_{i j}\right)^2\right]
\end{equation}

It is possible to get an EOM in the general case and it will have a form similar to scalar perturbations \eqref{EOMscal}, but provided that in our theory \eqref{lagrangian}:

\begin{equation}\label{GTFT}
    \mathcal{F}_T=\mathcal{G}_T=1,
\end{equation}

thus for any moment of time:

\begin{equation}
    c_T^2=1,
\end{equation}

and EOM takes standard form:

\begin{equation}\label{EOMten}
\ddot{h}_{i j}+3H\dot{h}_{i j}+\frac{k^2 c_T^2}{a^2}h_{i j}=0,
\end{equation}

where $\dot{ h }_{ij}$ is a derivative of $h_{ij}$ with respect to cosmic time $t$. Note that tensor perturbations contain two polarizations and can be decomposed as:

\begin{equation}
h_{i j}=h_{+}e_{i j}^{+}+h_{\times} e_{i j}^{\times}.
\end{equation}

Because of the factor in front of the action, canonically normalized initial conditions will have this form:

\begin{equation}
    h_{i j}(k,-\infty)=\frac{2}{a(t) (\mathcal{G}_T \mathcal{F}_T)^{\frac{1}{4}}}\frac{\exp{-i k \eta(t)}}{\sqrt{2c_T k}}=\frac{2}{a(t) }\frac{\exp{-i k \eta(t)}}{\sqrt{2 k}}.
\end{equation}

Thus we have equations of motion for scalars and tensors which can be solved for different momentums $k$.

\section{Lagrangian Reconstruction}
\label{3}

In this section we proceed to the construction of the model without initial singularity. Let us remind, that we consider a special case of the Horndeski theory in which the only non-zero functions in the action \eqref{action} are $G_2$ and $G_4$.
As we already took $G_4=\frac{1}{2}$, the result is GR with an additional scalar field. 
\subsection{Scale factor}

\begin{figure}[h]
    \centering
    \includegraphics[width=0.7\linewidth]{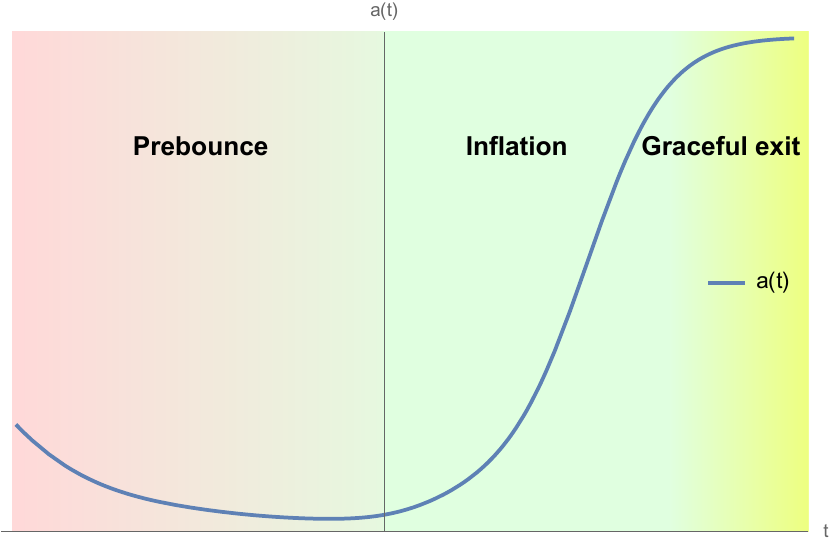}
    \caption{Principal form of scale factor.}
    \label{scalefac}
\end{figure}

The first step is to choose a scale factor that corresponds to an evolution without an initial singularity. We want to construct a scale factor that will describe a bounce universe with positive curvature, followed by an inflationary stage with subsequent slower expansion (graceful exit), allowing a transition to a hot stage. A principal form of such a scale factor is presented in Fig.\ref{scalefac}. 

We propose an analytical formulation for the scale factor, defined by four free parameters -- $\alpha$, $\beta$, $b$, and $n$ -- that influence its behavior and affect key cosmological parameters, including the scalar spectral index ($n_s$) and the tensor-to-scalar ratio ($r$). Additionally, there is one more free parameter, $a_0$ -- characteristic size of the universe, which we assume to be significantly larger than the Planck scale. However, variations in $a_0$ do not impact the power spectrum parameters. A comparison of scalar perturbation spectra for different values of $a_0$ is provided in the Appendix \ref{appendix:a1a10}.

\begin{equation}
   a(t)=\frac{10^{\alpha} t^2+a_0}{\exp{(bt)}+1}+\frac{10^{\beta}(t-t_b)^n+a_0}{\exp{(-bt)}+1}.
   \label{scalfac}
\end{equation}

It is noteworthy that within the scale factor \eqref{scalfac}, there is no transition to slower expansion. However, this transition can be seamlessly incorporated by employing an additional sigmoid function, facilitating a shift to phases such as kination, with the scale factor being proportional to $t^{\frac{1}{3}}$. Although the expression for the scale factor and the corresponding Lagrangian would be considerably extensive, the reconstruction technique remains the same. As our primary focus is on the parameters of the power spectra, we will concentrate solely on the initial two phases. The comprehensive expression of the scale factor is provided in the Appendix \ref{appendix:afull}. We also note that the contraction rate is chosen such that the curvature grows faster than the energy of the scalar field, in our case $a_{con}(t) \propto t^2$ for simplicity.

\subsection{Why do we need kinetic terms and curvature?}

First, it is important to note that for the chosen scale factor, during the transition from the prebounce stage to inflation, there is a period when $\dot{H} > 0$. In a flat universe, this requires the presence of NEC-violating matter \cite{Rubakov_2014}, but one way to bypass this is by considering positive spatial curvature, which compensates for $\dot{H}$ in the bounce stage.

This can be straightforwardly observed from the form of the Friedmann equations for a classical scalar field:

\begin{equation}
\begin{aligned}
& 3 \dot{H}+3 H^2=V(\phi)-\dot{\phi}^2 \\
& 3 H^2+3 \frac{1}{a^2}=\frac{1}{2} \dot{\phi}^2+V(\phi)
\end{aligned}
\end{equation}

thus,

\begin{equation}
\dot{\phi}^2(t)=2\left(\frac{1}{a^2}-\dot{H}\right)
\end{equation}

without the $\frac{1}{a^2}$ term there is no real stable solution. Similarly, it can be observed that potential term $V(\phi)$ is growing up before bounce stage, and then slowly decreases on inflation stage: 

\begin{equation}
V(\phi)=\dot{H}+3H^2+\frac{2}{a^2}
\end{equation}

One might question the necessity of considering non-trivial kinetic terms when constructing the Lagrangian for such a theory. According to \cite{Ellis:1990wsa}, the potential is enough for a wide range of desirable behaviors of the scale factor. The key feature of our approach lies in the absence of the need to solve differential equations to obtain the Lagrangian. Instead, we derive the Lagrangian by solving algebraic equations, allowing us to find its exact analytical form without a necessity to use numerical approaches. This method requires at least two functions in the ansatz: the potential $V(\phi)$ and a non-trivial kinetic term.

\subsection{Reconstruction}

We are now prepared to reconstruct the Lagrangian. Let us consider this ansatz:

\begin{equation}
G_2(\phi, X)=g_0(\phi)+g_1(\phi) X+g_2(\phi) X^2.
\end{equation}

As previously introduced $G_4=\frac{1}{2}$, so $\mathcal{F_T}=\mathcal{G_T}=1$, and we require $\mathcal{F_S}=\mathcal{G_S}$, which leads to

\begin{equation}
    c^2_S=c^2_T=1.
\end{equation}

It is obvious from \eqref{cs} that \( c^2_s = 1 \) results in \( G_{2XX}=2g_2 = 0 \), hence our ansatz simplifies to \( G_2 = g_0(\phi) + g_1(\phi) X \). We substitute it into equations of motion \eqref{eom1}, \eqref{eom2} and solve the system of algebraic equations to obtain expressions for the functions in the Lagrangian. Here we consider $\phi(t)=t, X=\frac{1}{2}$ on shell without loss of generality.

 Note that by redefining the fields, our ansatz can be reduced to a conventional scalar field with a non-trivial potential and a standard kinetic term. Thus, there is no necessity to consider k-essence theory; however, this freedom in choosing the function for the kinetic term allows for an easy reconstruction of the Lagrangian.

The explicit forms of the functions \(g_i\) for this ansatz, with arbitrary coefficients in the scale factor \eqref{scalfac}, are presented in Appendix~\ref{appendix:g1g2}. However, the explicit expressions for \(g_i\) corresponding to the scale factor with a graceful exit \eqref{fulscalfac} are exceedingly lengthy, even for the appendix. Instead, we provide the characteristic behavior of \(g_i\) associated with \eqref{fulscalfac} in Fig.~\ref{qw}. It is worth noting that the asymptotic behavior at \(t \to +\infty\) is such that \(g_0 \propto -\frac{1}{t^4}\) and \(g_1 \propto \frac{1}{t^2}\). We demonstrate that both functions, representing the potential and kinetic terms, converge to zero at \(t \to \pm\infty\), corresponding to a regime dominated by General Relativity with a massless scalar field in the future.

\begin{figure}\label{qw}
    \centering    \includegraphics[width=0.7\linewidth]{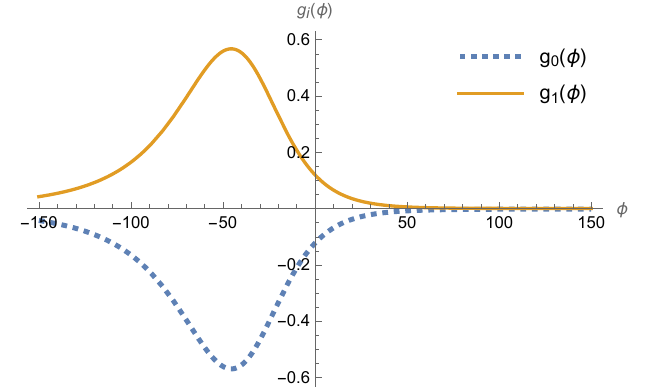}
    \caption{Components of $G_2$ function throughout the whole evolution.}
    \label{qw}
\end{figure}

\subsection{No fine tuning}

It is important to note that the solution obtained for the scale factor is not fine-tuned. Consequently, when attempting to resolve the equations of motion derived from the reconstructed Lagrangian under alternative initial conditions, one obtain solutions of the same nature.

Let us examine it. Using the obtained expressions for \( g_i \), we can analyze the solutions of the Friedmann equations \eqref{eom1} and \eqref{eom2}:

\begin{equation}\label{stbl1}
     \frac{1}{2}g_1(\phi) \dot{\phi}^2-g_0(\phi)-3(\frac{\dot{a}}{a})^2-\frac{3}{a^2}=0,
\end{equation}

\begin{equation}\label{stbl2}
    \frac{1}{2}g_1(\phi) \dot{\phi}^2+g_0(\phi)+(\frac{\dot{a}}{a})^2+2\frac{\ddot{a}}{a}+\frac{1}{a^2}=0.
\end{equation}

\begin{figure}
    \centering
    \includegraphics[width=0.7\linewidth]{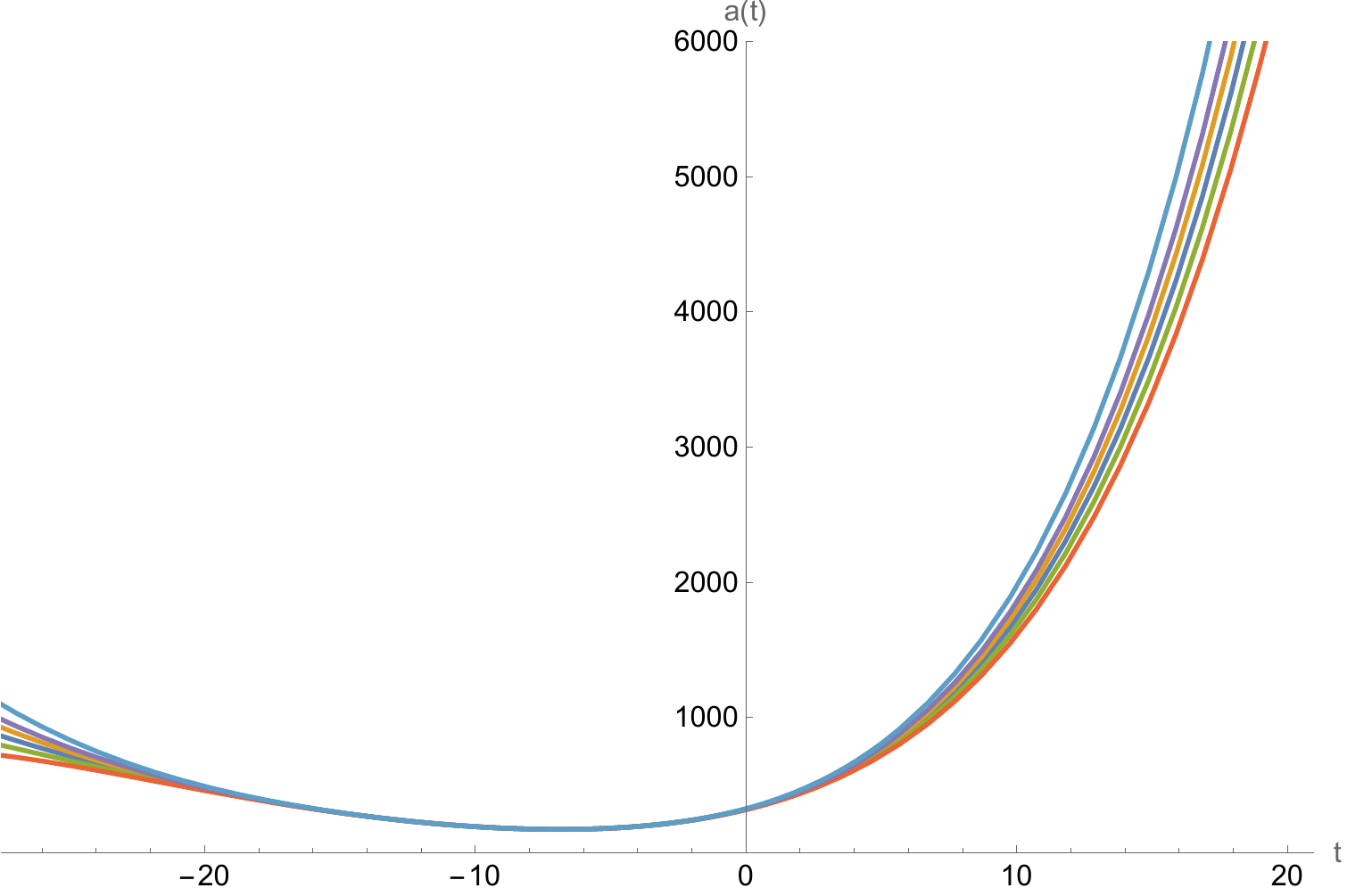}
    \caption{Numerical solution of \eqref{stbl1} and \eqref{stbl2} for slightly different initial conditions for $a(t)$.}
    \label{a in con}
\end{figure}

If we slightly change the initial conditions for the scale factor in the past we obtain a slightly different solution, nevertheless with the same behavior, which indicates that the solution is stable with respect to the initial conditions. Numerical solutions for slightly different initial conditions are represented in Fig. \ref{a in con}. It is essential that despite the different initial conditions this has almost no effect on the inflation period and thus on the observable parameters. 

It is worth noting that the absence of fine-tuning is a key distinction from models that exhibit an Einstein universe asymptotic behavior in the past \cite{2Ellis_2003, Ellis_2003}. We have also conducted numerical analyses of models with initial conditions corresponding to an Einstein universe, that is, a universe with constant curvature. However, such solutions are unstable with respect to changes in initial conditions: even a small perturbation leads to a fundamental alteration in the behavior of the scale factor, resulting either in a future singularity or in a bounce. This is precisely why the bounce solution is of particular interest, as it represents a model with the most natural initial conditions, deviations from which do not alter the characteristic behavior of the scale factor.

\subsection{Anisotropy growth on the contraction and BKL instability}

One of the key issues to address when constructing bouncing cosmologies is the Belinskii–Khalatnikov–Lifshitz (BKL) phenomenon \cite{Belinsky:1970ew}, which may lead to strong inhomogeneity and anisotropy of space by the end of the contraction stage. This anisotropy is diluted during the subsequent inflationary phase. However, it requires one to consider anisotropic generalization of the Friedmann equations, which may spoil the desired behavior around the bounce itself.

One possible solution to the BKL instability in GR is to introduce a matter component with a super-stiff equation of state, $p \geq \rho$, during the contracting phase, where $\rho$ and $p$ denote energy density and effective pressure, respectively. For instance, ekpyrotic scenario \cite{Khoury_2001} or massless scalar field with the equation of state $p=\rho$  
dominating during the contraction epoch  \cite{Khalatnikov_2003}.

 In our model, we assume that the contraction phase lasted for a finite duration, during which the anisotropy grew but remained small. Specifically, we require that its contribution to the total energy density at the bounce did not exceed the contribution from spatial curvature. This ensures that the bounce is curvature-driven, rather than anisotropy-driven (although the latter scenario might also be interesting for further study). Thus, while our model does not require fine-tuning of the Friedmann equations themselves, we implicitly assume the existence of an earlier phase (such as in a cyclic cosmology) that suppressed the initial anisotropy to very small values. As a result, although anisotropies increased during contraction, their contribution remained subdominant compared to curvature, and the FRW approximation held throughout.

One possible way to impose strict conditions on anisotropies before the contraction phase — whether in a cyclic or some other type of universe — is to consider the specific behavior of tensor and scalar perturbations at earlier stages. Returning to the perturbation equations \eqref{EOMscal} and \eqref{EOMten}, one can derive conditions under which BKL-like modes are absent \cite{Ageeva_2023}. In these equations, we consider the limit of infinitely large wavelengths ($k \rightarrow 0$). This limit corresponds to selecting a preferred spatial direction associated with anisotropy. To suppress the BKL instability, we require that these perturbation equations do not admit growing solutions for either the tensor or the scalar modes:

\begin{equation} \frac{1}{a^3 \mathcal{G}_T} \frac{d}{d t}\left(a^3 \mathcal{G}_T \dot{h}{i j}\right) = 0, \end{equation}

\begin{equation} \frac{1}{a^3 \mathcal{G}_S} \frac{d}{d t}\left(a^3 \mathcal{G}_S \dot{\zeta}\right) = 0. \end{equation}

Thus, an appropriate choice of the functions $\mathcal{G}_S$ and $\mathcal{G}_T$ at earlier stages (although going beyond the scope of the present work) 
provides a possible way to prepare a very smooth initial state before the onset of contraction.

It is worth noting that in the already mentioned Einstein static universe scenario, where the universe evolves through an unstable quasi-static state with fixed spatial curvature, this problem is entirely absent because the constant scale factor prevents anisotropies from growing. In such models, one typically needs to fine-tune the initial conditions for the scale factor and the background scalar field, but no strong constraints on anisotropy are required.

\section{Observables }
\label{4}

Now that we have established the Lagrangian, we proceed to the analysis of the Power spectrum. Equations \eqref{EOMscal} and \eqref{EOMten} are solved numerically, employing vacuum initial conditions for a range of momentum values $k$. To calculate the spectrum we find the value of the solutions at the time of inflationary expansion, i.e. before the graceful exit. Technically, this means that we solve the equation for a Fourier mode in some time interval, setting the initial conditions at the moment long before the bounce,  when $H$ is small and the vacuum initial conditions can be employed, and find its value at the end of this interval before the graceful exit.

The scalar power spectrum $\mathcal{P}_{\zeta} (k) $ is obtained by evaluating the two-point correlation function of $\zeta$ in Fourier space:

\begin{equation}
\mathcal{P}_{\zeta} (k)=\frac{k^3}{2 \pi^2}\left|\zeta_k\right|^2.
\end{equation}

Mathematically, we calculate the square of the module for the value of $\zeta$ at a specific point related to inflation. Similar reasoning can be done for tensor modes, but it is necessary to keep in mind two polarization states:

\begin{equation}
\mathcal{P}_h^{(s)}(k)=\frac{k^3}{2 \pi^2}\left|h_s(k)\right|^2, 
\end{equation}

where $s$ represents each polarization state, $s=+, \times$. Since there are two independent polarization states, the total power spectrum is the sum of the power spectra of the individual polarization, which are the same in our case:

\begin{equation}
\mathcal{P}_h(k)=\mathcal{P}_h^{(+)}(k)+\mathcal{P}_h^{(\times)}(k)=2 \mathcal{P}_h^s(k)
\end{equation}

Then we calculate the spectral index $n_s$, which is defined as:

\begin{equation}
\mathcal{P}_{\zeta}(k)=A_S\left(\frac{k}{k_0}\right)^{n_S-1}, 
\end{equation}

and for tensor modes:

\begin{equation}
\mathcal{P}_{h}(k)=A_T\left(\frac{k}{k_0}\right)^{n_T}, 
\end{equation}

where $k_0$ corresponds to
the comoving wave number of the perturbation mode which exits horizon around the beginning of inflation.

The ratio between tensor and scalar perturbations is characterized by the tensor-to-scalar ratio $r$, defined as:

\begin{equation}
r=\left.\frac{\mathcal{P}_h(k)}{\mathcal{P}_{\zeta}(k)}\right|_{k=k_0}.
\end{equation}

It is important to note that by varying the parameters of the scale factor it is possible to achieve almost any values of $r$ and $n_s$, so we find several sets that satisfy both  stability conditions \eqref{stable2}  and current restrictions \cite{Tristram_2021} for $r$ and $n_s$.

\begin{figure}[h!]
    \centering
    \includegraphics[width=0.7\linewidth]{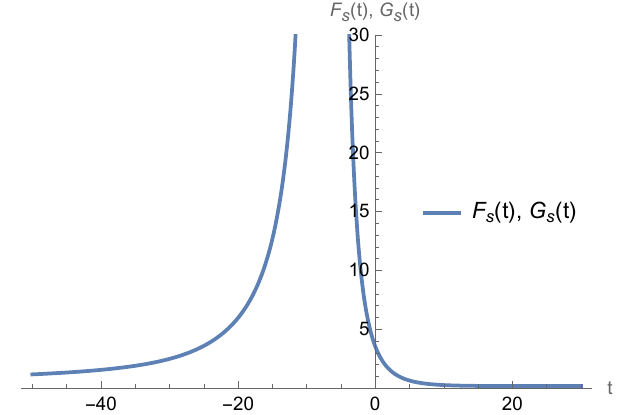}
    \caption{Stability conditions for $\mathcal{G_S}, \mathcal{F_S}$.}
    \label{GS}
\end{figure}

The characteristic behavior of the functions $\mathcal{G}_S$ and $\mathcal{F}_S$ is shown in the Fig.~\ref{GS}. While they may appear to approach zero, they actually converge to a constant determined by the growth rate of the scale factor. Let us note that near to the bounce the so-called $\gamma$-crossing occurs in our model. It was discussed in details in \cite{Mironov:2018oec}, where it was shown to be an artifact of the gauge fix, so despite the singularity in the coefficients of second order action, the solutions for perturbations remain smooth. We emphasize that the scalar speed of sound is unity everywhere.

\begin{figure}[h]
    \centering
    \includegraphics[width=0.7\linewidth]{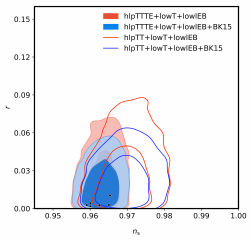}
    \caption{Table 1 and marginalized joint 68\,\% and 95\,\% CL regions for $n_{\rm s}$ and $r_{0.05}$ from Planck alone (hlp+lowT+lolEB) and in combination with BK15. The solid lines correspond to hlp TT+lowT+lolEB, while the filled regions include TE and correspond to hlp TTTE+lowT+lolEB \cite{Tristram_2021}.}
    \label{ns_r}
\end{figure}

We show several sets of parameters that are consistent with the observational data\footnote{
Other spectral parameters, such as the running of the spectral index and the tensor spectral index $n_T$, can also be calculated. However, we do not provide specific values for these parameters, since there are currently no stringent constraints on them.} in Table \ref{tab:parameters}. Let us note, that the presented points are merely an examples, as the model admits wide range of experimental signatures for different parameters.  Figure \ref{ns_r} is a combined plot of the current observational data and the points from Table \ref{tab:parameters}.

\begin{table}[h!]
\centering
\renewcommand{\arraystretch}{1.5}  
\setlength{\tabcolsep}{12pt} 
\caption{Examples of parameters satisfying the current observational data.}
\begin{tabular}{|c|c|c|c|c|c|c|}
\toprule
$\alpha$ & $\beta$ & $b$ & $n$ & $t_b$ & $n_s$ & $r$ \\
\midrule
-3.57 & -162.56 & 0.023 & 50 & -1850 & 0.959 & 0.0023 \\
-2.37 & -168.00 & 0.020 & 52 & -1790 & 0.960 & 0.0018 \\
-2.69 & -168.00 & 0.060 & 52 & -1750 & 0.960 & 0.0026 \\
-2.69 & -180.00 & 0.080 & 56 & -1675 & 0.962 & 0.0023 \\
-2.69 & -190.00 & 0.010 & 60 & -1525 & 0.965 & 0.0024 \\
-2.22 & -189.52 & 0.040 & 60 & -1495 & 0.966 & 0.0108 \\
\bottomrule
\end{tabular}
\label{tab:parameters}
\end{table}

\section{Conclusions}\label{sec conclusions}
\label{5}

We have constructed an inflationary model in a closed Universe without an initial singularity, which remains
stable without the necessity to incorporate NEC-violated matter. The model is free from pathologies such as gradient or tachyonic instabilities. A key distinction from other models is the explicit form of the scale factor and possibility to reconstruct the Lagrangian of a scalar field in the exact form.

Furthermore, it has been demonstrated that a broad range of observational data can be obtained by varying the
parameters of the scale factor. We have presented several examples that satisfy the current constraints on $n_s$ and $r$. However, it is important to note that the future observations may allow for the identification of new parameters, which can be easily calculated within this theoretical framework.

In summary, we have described a stable cosmological model without an initial singularity, capable of predicting different sets of observational data.


\section*{Acknowledgements}
The author is deeply grateful to Sergey Mironov for his original idea, valuable remarks throughout the work and careful reading of the manuscript.

The work on Section \ref{3} of this paper 
has been supported by the Foundation for the Advancement of Theoretical
Physics and Mathematics "BASIS", the work on Section \ref{4} has been supported by
Russian Science Foundation grant № 24-72-10110, https://rscf.ru/project/24-72-10110/.

\newpage
\appendix

\section{Full form of the scale factor}
\label{appendix:afull}

Here is the full form of the scale factor with exit to the kination phase, dominated by a massless scalar field:
\begin{equation}
   a(t)=\left(\frac{10^{\alpha} x^2+a_0}{\exp{(bx)}+1}+\frac{10^{\beta}(x-x_b)^n+a_0}{\exp{(-bx)}+1}\right) \frac{1}{\exp{(dx-N)}+1}+\frac{10^\gamma (x^2-x_k^2)^{\frac{1}{6}}}{\exp{(-dx+N)}+1}.
   \label{fulscalfac}
\end{equation}
Note that $N$ and $d$ are responsible for inflation stage duration (number of e-folds).

\section{Lagrangian functions}\label{appendix:g1g2}

In this Appendix, we collect the expressions for the Lagrangian functions $g_i$ in the following ansatz: 
\begin{equation}
\begin{aligned}
&G_4=\frac{1}{2}, \\
&G_2=g_0+g_1 X.
\end{aligned}
\end{equation}

Note that this functions represent only first two stages in scale factor \eqref{fulscalfac}: contraction and inflation, because expressions for full Lagrangian are huge even for Appendix.
\begin{equation}
\begin{gathered}
g_0(\phi)=\left(( e ^ { b \phi ( t ) } + 1 ) ^ { 2 } \left(-\frac{1}{e^{b \phi(t)}+1}\left(\frac{2^{\alpha+1} \times 5^\alpha+10^\beta(n-1) n e^{b \phi(t)}(\phi(t)-x_b)^{n-2}}{e^{b \phi(t)}+1}+\right.\right.\right. \\
+\left(b e ^ { b \phi ( t ) } \left(10^\beta(\phi(t)-x_b)^n\left(e^{b \phi(t)}(b x_b-b \phi(t)+2 n)-b x_b+b \phi(t)+2 n\right)+\right.\right. \\
\\
+10^\alpha \phi(t)+\left.\left.\left.\left(-b \phi(t)+e^{b \phi(t)}(b \phi(t)-4)-4\right)(\phi(t)-x_b)\right)\right) /\left(\left(e^{b \phi(t)}+1\right)^3(\phi(t)-x_b)\right)\right) \\
\left(a_0 e^{b \phi(t)}+a_0+10^\beta e^{b \phi(t)}(\phi(t)-x_b)^n+10^\alpha \phi(t)^2\right)-\frac{1}{\left(e^{b \phi(t)}+1\right)^4} \\
\left.\left.2\left(10^\beta e^{b \phi(t)}(\phi(t)-x_b)^{n-1}\left(-b x_b+n e^{b \phi(t)}+b \phi(t)+n\right)-10^\alpha \phi(t)\left(e^{b \phi(t)}(b \phi(t)-2)-2\right)\right)^2-2\right)\right) \times \\
\\
\times\frac{1}{\left(a_0 e^{b \phi(t)}+a_0+10^\beta e^{b \phi(t)}(\phi(t)-x_b)^n+10^\alpha \phi(t)^2\right)^2}
\end{gathered}
\end{equation}

\begin{equation}
\begin{gathered}
g_1(\phi)=\left(2 \left(1 0 ^ { \beta } e ^ { b \phi ( t ) } \left(10^\beta e^{2 b \phi(t)}\left(b^2 x_b^2+2 n+2 n \cosh (b \phi(t))\right)(\phi(t)-x_b)^n-\left(1+e^{b \phi(t)}\right)\left(a_0(n-1) n\left(1+e^{b \phi(t)}\right)^2-\right.\right.\right.\right. \\
\left.\left.2 a_0 b x_b n\left(1+e^{b \phi(t)}\right)+\left(2^{\alpha+1} \times 5^\alpha\left(1+e^{b \phi(t)}\right)-a_0 b^2\left(-1+e^{b \phi(t)}\right)\right) x_b^2\right)\right)(\phi(t)-x_b)^n+ \\
100^\alpha b^2 e^{b \phi(t)} \phi(t)^6-2^{2 \alpha+1} \times 25^\alpha b^2 e^{b \phi(t)} x_b \phi(t)^5+\left(1+e^{b \phi(t)}\right)^3\left(-2^{\alpha+1} \times 5^\alpha a_0+e^{b \phi(t)}+1\right) x_b^2+ \\
2^{\alpha+1} \times 5^\alpha \phi(t)^3\left(2^{\beta+1} \times 5^\beta b e^{2 b \phi(t)}(-n+(b x_b-n+2) \cosh (b \phi(t))+2)(\phi(t)-x_b)^n+\right. \\
\left.\left(1+e^{b \phi(t)}\right)\left(a_0 b e^{b \phi(t)}\left(-b x_b+e^{b \phi(t)}(b x_b+2)+2\right)-2^{\alpha+1} \times 5^\alpha\left(1+e^{b \phi(t)}\right) x_b\right)\right)+ \\
\phi(t)^2\left(1 0 ^ { \beta } e ^ { b \phi ( t ) } \left(a_0 e^{2 b \phi(t)} b^2-a_0 b^2-2^{\alpha+1} \times 5^\alpha e^{b \phi(t)}(-2 b x_b(n-4)+(n-5) n+\right.\right. \\
\left.\left.\left(n^2-2 b x_b n-5 n+b x_b(b x_b+8)+2\right) \cosh (b \phi(t))+2\right)\right)(\phi(t)-x_b)^n+100^\beta b^2 \\
e^{3 b \phi(t)}(\phi(t)-x_b)^{2 n}+\left(1+e^{b \phi(t)}\right)\left(\left(1+e^{b \phi(t)}\right)\left(2^{2 \alpha+1} \times 25^\alpha x_b^2+2 e^{b \phi(t)}+e^{2 b \phi(t)}+1\right)-\right. \\
\left.\left.10^\alpha a_0\left(e^{b \phi(t)}\left(b x_b(8-b x_b)+e^{b \phi(t)}(b x_b(b x_b+8)+2)+4\right)+2\right)\right)\right)+ \\
2 \phi(t)\left(1 0 ^ { \beta } e ^ { b \phi ( t ) } ( 1 + e ^ { b \phi ( t ) } ) \left(-a_0\left(-1+e^{b \phi(t)}\right) x_b b^2+2^{\alpha+1} \times 5^\alpha\left(1+e^{b \phi(t)}\right) x_b^2 b-a_0\left(1+e^{b \phi(t)}\right)\right.\right. \\
\left.n b-2^{\alpha+1} \times 5^\alpha\left(1+e^{b \phi(t)}\right) x_b(n-1)\right)(\phi(t)-x_b)^n-100^\beta b^2 e^{3 b \phi(t)} x_b(\phi(t)-x_b)^{2 n}+ \\
\left.\left(1+e^{b \phi(t)}\right)^2 x_b\left(2^{\alpha+1} \times 5^\alpha a_0\left(e^{b \phi(t)}(b x_b+1)+1\right)-\left(1+e^{b \phi(t)}\right)^2\right)\right)+10^\alpha \phi(t)^4 \\
\left(\left(1+e^{b \phi(t)}\right)^2(x_b-\phi(t))^2\left(10^\beta e^{b \phi(t)}(\phi(t)-x_b)^n+a_0 e^{b \phi(t)}+10^\alpha \phi(t)^2+a_0\right)^2\right)
\end{gathered}
\end{equation}

\section{Comparison of power spectra for different universe sizes}\label{appendix:a1a10}
As we have already noted, the characteristic size of the universe does not affect on power spectra observables. Here is the representation of scalar power spectra for all the same parameters in scale factors, except of $a_0$.

 \begin{figure}[h!]
    \centering
    \includegraphics[width=0.68\linewidth]{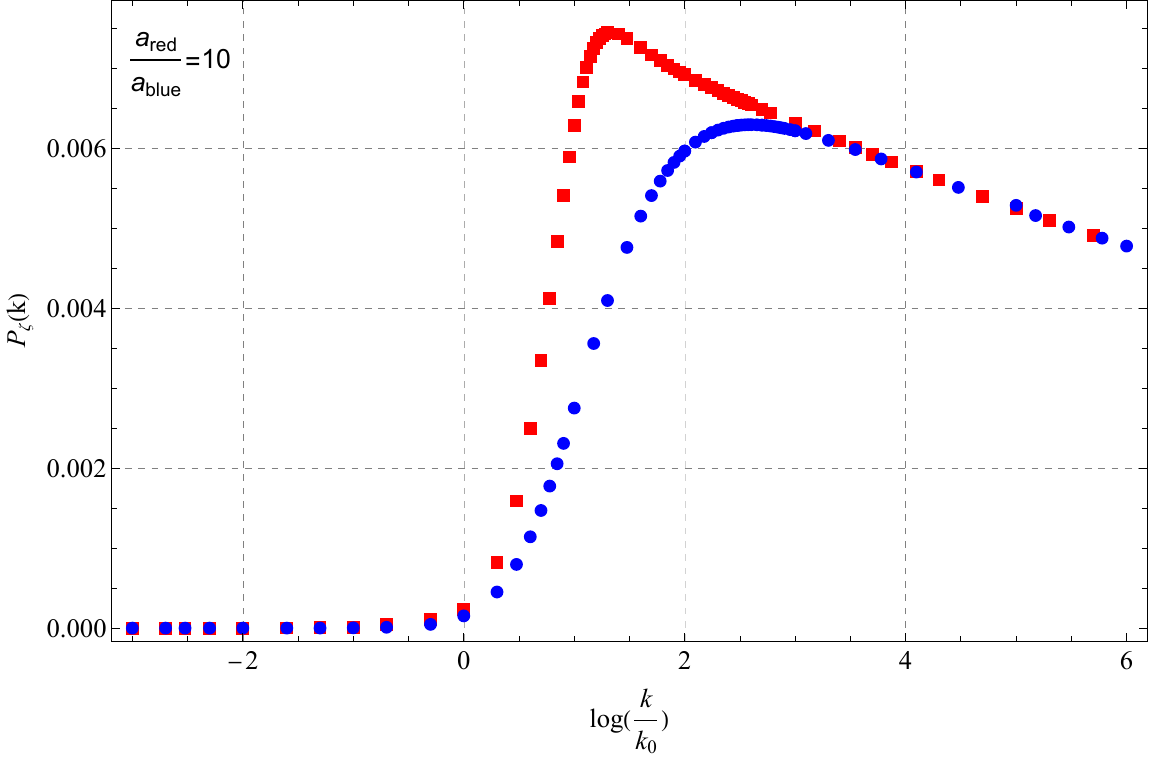}
    \caption{Scalar power spectra for different $a_0$. Red squares correspond to a scale factor 10 times larger than the scale factor corresponding to blue dots.}
    \label{fig:enter-label}
\end{figure}

\newpage

\bibliographystyle{IEEEtran}

\bibliography{main}

\end{document}